\documentclass[10pt, numberedappendix]{iopart}
\usepackage{graphicx}
\usepackage{amsfonts}
\usepackage{amssymb}
\usepackage{iopams}
\usepackage{threeparttable}
\begin{document}
\title[High-order Dy multipoles in antiferromagnetic DyB$_2$C$_2$]{Triakontadipole and high-order dysprosium multipoles in the antiferromagnetic phase of DyB$_2$C$_2$}
\date{\today}
\author{A~J~Princep$^{1,2}$, A~M~Mulders$^{1,2,3}$, U~Staub$^4$, V~Scagnoli$^4$, T~Nakamura$^5$, A~Kikkawa$^6$, S~W~Lovesey$^{7,8}$, E~Balcar$^9$}
\address{$^1$University of New South Wales, Canberra Campus, ACT, 2600, Australia}
\address{$^2$Department of Imaging and Applied Physics, Curtin University, Perth, WA, 6845, Australia}
\address{$^3$Bragg Institute, ANSTO, Lucas Heights, NSW, 2234, Australia}
\address{$^4$Swiss Light Source, Paul Scherrer Institut, CH-5232 Villigen PSI, Switzerland}
\address{$^5$JASRI/SPring-8, Sayo, Hyogo 679-5198, Japan}
\address{$^6$RIKEN, 2-1, Hirosawa, Wako, Saitama 351-0198, Japan}
\address{$^7$ISIS Facility, Harwell Science and Innovation Campus, Oxfordshire OX11 0QX, U.K}
\address{$^8$Diamond Light Source Ltd., Oxfordshire OX11 0DE, U.K.}
\address{$^9$Vienna University of Technology, Atominstitut, Wiedner Hauptstrasse 8-10, 1040 Vienna, Austria}
\ead{andrew.princep@student.adfa.edu.au}
\begin{abstract}
Resonant soft x-ray Bragg diffraction at the Dy M$_{4,5}$ edges has been used to study Dy multipoles in the combined magnetic and orbitally ordered phase of DyB$_2$C$_2$. The analysis incorporates both the intra-atomic magnetic and quadrupolar interactions between the 3d core and 4f valence shells. Additionally, we introduce to the formalism the interference of magnetic and nonmagnetic oscillators. This allows a determination of the higher order multipole moments of rank 1 (dipole) to 6 (hexacontatetrapole). The strength of the Dy 4f multipole moments have been estimated at being between 7 and 78\% of the quadrupolar moment.  
\end{abstract}
\section{Introduction}

Resonant x-ray Bragg diffraction (RXD), which is Bragg diffraction where the energy of the incident x-rays are tuned to a relevant absorption edge, has established itself as a powerful tool in modern solid state physics to investigate magnetic, orbital and charge ordering phenomena associated with electronic degrees of freedom. The data obtained from an RXD experiment can be interpreted using a model based on purely atomic quantities, neglecting e.g. band and hybridization effects. Models based on atomic quantities have the advantage of being suited to the analysis of both x-ray and neutron diffraction experiments, as well as absorption, NMR, EPR, muon and Mossbauer spectroscopies \cite{electronicproperties}. The localized nature of rare earth 4$f$ electrons makes them particularly suited to such analysis which conveniently relies upon quantities such as atomic tensors and operator equivalents with the many simplifications that these bring. In particular, any atomic observable may be decomposed into its $(2l+1)^2$ multipole operator components $T_q^x$ (with $x=0,...,2l$ and $q=-x,...,x$, where $l$ is the orbital angular momentum quantum number corresponding to the electronic shell in question), and any atomic state may be represented by the set of nonvanishing multipole components  $\langle T_q^x \rangle $ \cite{paradigm}. $\langle \ldots \rangle$ denotes the expectation value of the enclosed operator on the ground state of the electrons. Multipoles $\langle T_q^x \rangle$ are parity even. Additionally, they have the definite time signature $(-1)^x$, and we refer to time-odd multipoles as magnetic and time-even multipoles as charge. In Rare Earth compounds, the 4$f$ electrons are not strongly coupled to the lattice and they may retain their degrees of freedom down to very low temperatures, and although ordered states of magnetic dipoles are relatively common, multipoles of rank 2 (quadrupole) or higher may order under favorable conditions \cite{multipoleorders}. Orbital ordering is usually labeled  as simply antiferroquadrupolar (AFQ) or ferroquadrupolar (FQ) since higher-order multipoles are not well established by observations, despite the fact that for 4$f$ electrons ($l = 3$) multipoles up to rank $6$ can be present. In materials with a phase attributed to such ``hidden'' ordering, exactly which multipoles dominate the transition to the ordered phase remains a rather controversial topic,  as exemplified by the compounds Ce$_x$La$_{1-x}$B$_6$ \cite{cerium}, NpO$_2$ \cite{neptunium} and particularly URu$_2$Si$_2$ \cite{uranium}. RXD is particularly useful in characterizing such orders, and applications of this technique have produced both direct and indirect evidence of high-order multipole moments in DyB$_2$C$_2$ \cite{tanaka, hirota}, Ce$_x$La$_{1-x}$B$_6$ \cite{celab}, and NpO$_2$ \cite{Np}.

DyB$_2$C$_2$  has the highest known AFQ ordering temperature of any rare earth systemT$_{Q}$ at 24.7 K and as such it has attracted much attention, as many other compounds where AFQ or FQ ordering is postulated order at temperatures of 5 K or less \cite{paradigm}. At room temperature, DyB$_2$C$_2$ exhibits the tetragonal structure P4$/{mbm}$ and undergoes a transition with small alternating shifts of pairs of B and C atoms along c at T$_Q$ \cite{dystructure} which reduces the symmetry to P4$_{2}/{mnm}$ \cite{dysymmetry}. Below T$_{\textrm{M}}$ = 15.3 K, antiferromagnetic (AFM) order with four distinct magnetic sublattices is observed, and the moment orientations are postulated to arise from the underlying orbital interaction \cite{dymagnetism}. Additionally the crystal evolves a monoclinic lattice strain at T$_{\textrm{N}}$ \cite{ultrasound}. While the properties of combined AFM/ AFQ phase (Phase IV) of the similar compound HoB$_2$C$_2$ are postulated to arise from an octupole order parameter \cite{ultrasound}, the case for DyB$_2$C$_2$ is not as clear cut due to the presence of the aforementioned lattice strain, and the broader variation in its elastic properties at the AFM and AFM+AFQ transitions. 

Due to the less stringent experimental constraints, most RXD studies of rare earth materials using x-rays probe the L$_{2,3}$ edges via an E1 dipole transition (2$p$ $\rightarrow$ 5$d$) or an E2 quadrupole transition (2$p$ $\rightarrow$ 4$f$). In most cases the information obtained from such a measurement is relatively indirect in the case of an E1 resonance, and with relatively low intensity in the case of an E2 resonance. Soft Resonant X-ray Bragg Diffraction (SRXD) at the Rare Earth M$_{4,5}$ edges probes the empty 4$f$ valence states directly by exciting a dipole resonance (3$d$ $\rightarrow$ 4$f$). SRXD experiments at the Dy M$_{4,5}$ edges have shown that the Coulomb (intra-atomic quadrupole) interaction between the 3d and 4f shells is significant \cite{us}. The ordered quadrupole moment of the 4f shell in the intermediate state reshapes the observed 3d$_{5/2}$ and 3d$_{3/2}$ core hole charge density and leads to a splitting of the 3$d$ energy levels. This splitting causes interference between different pathways of the scattering amplitude, allowing the observation of high-order multipole moments of the 4f shell that have rank 3, 4, 5 and rank 6: respectively the  octupole, hexadecapole, triakontadipole, and the hexacontatetrapole. Quantifying these higher order multipoles allows the ground state of dysprosium to be expressed in terms of the associated state multipoles, and accordingly a better understanding of its electronic structure can be obtained.

In the present work, we probe the combined AFM / AFQ phase of DyB$_2$C$_2$  using the same core hole splitting formalism previously applied to the AFQ phase  \cite{us}, with the structure factor extended to include magnetism. Phenomenological changes are introduced to mimic differences between magnetic resonances and nonmagnetic ones. The presence of magnetic ordering allows the measurement of odd rank multipoles (1,3,5) associated with the magnetic order, and it is observed that their presence changes the even rank $\langle T_q^x \rangle$. 

\section{Experimental Results}

A DyB$_2$C$_2$ single crystal was grown by the Czochralski method using an arc-furnace with four electrodes and cut with $\left( 0 0 1 \right)$ perpendicular to the sample surface. Subsequently, it was
polished and aligned with 26$^{\circ}$, 56$^{\circ}$ or 86$^{\circ}$ azimuthal angle. Zero degree azimuth corresponds to alignment of the b axis in the scattering plane. The orbital ordering $\left( 0 0 \frac{1}{2} \right)$ reflection was recorded (without polarization analysis \cite{jsynchrad}) at the Dy M$_{4,5}$ edges of DyB$_2$C$_2$ at the RESOXS end-station of the SIM beam-line at the Swiss Light Source. The Dy M$_{4,5}$ absorption edges were characterized with fluorescence yield (FY) and electron yield (EY) at RESOXS and the BL25SU beam line at SPring-8, respectively.

\begin{figure}
\begin{centering}
\includegraphics[width=\textwidth]{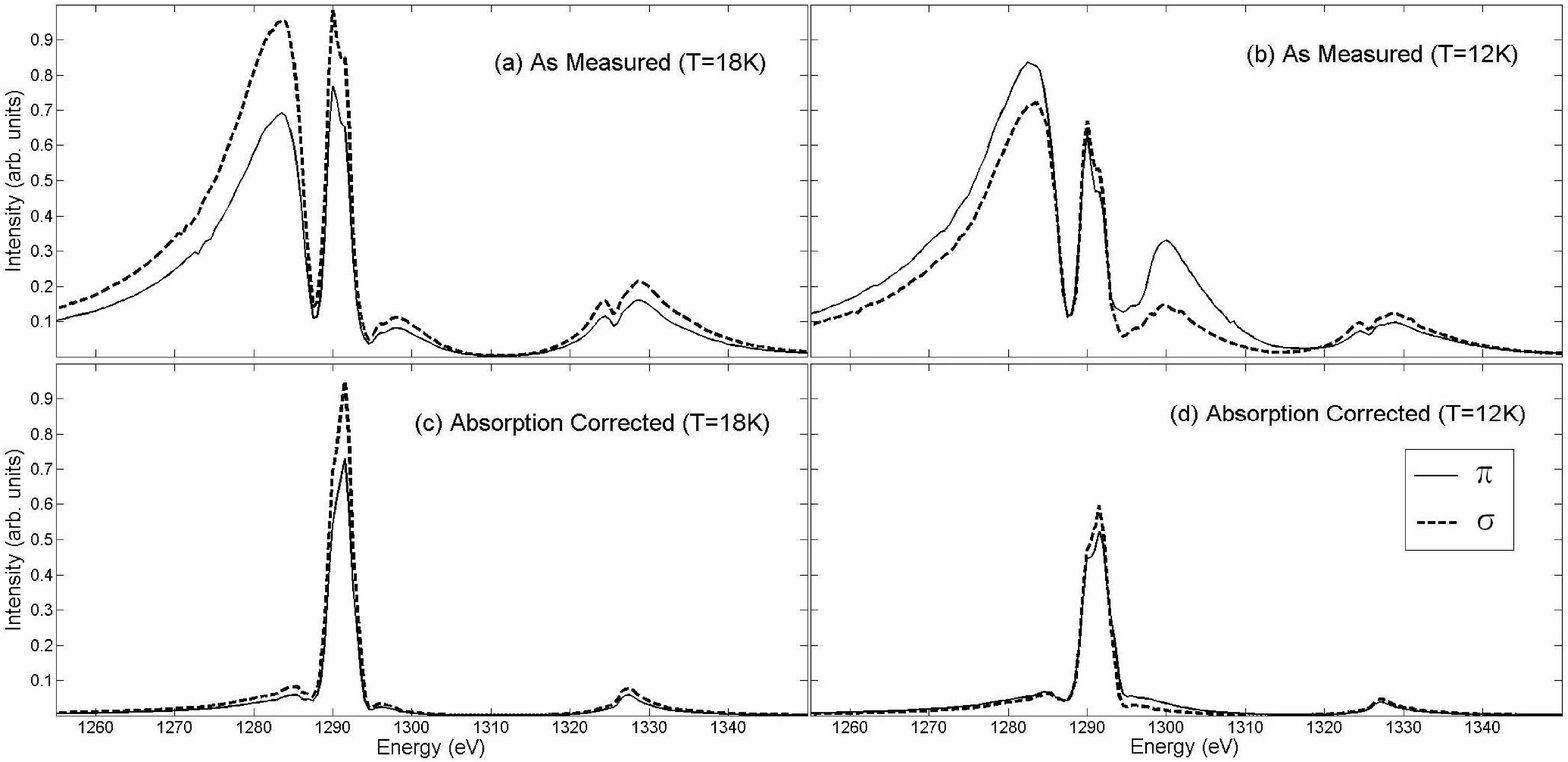}
\caption{\label{rawdata} (a) and (b) Diffracted intensity of the $(00\frac{1}{2})$ reflection in DyB$_2$C$_2$ taken with incident $\sigma$ (vertical) and $\pi$ (horizontal) polarization at 18K and 12K respectively. (c) and (d) diffracted intensity of (a) and (b) corrected for absorption effects.  Note: All data is normalized to the central peak at the M$_5$ edge in the $\sigma$ channel, and the ratio between $\sigma$ and $\pi$ channels is explicitly maintained.}
\end{centering}
\end{figure}

The recorded energy profile of the space group-forbidden $\left( 0 0 \frac{1}{2} \right)$ reflection of DyB$_2$C$_2$ in the pure AFQ phase \cite{us} and the AFM + AFQ phases are shown in Figure~\ref{rawdata} (a) and (b) respectively. The integrated intensity was corrected for absorption, and the corrected curves are shown in Figure~\ref{rawdata} (c) and (d). We draw particular attention to the qualitative difference between the high and low temperature spectra, due to the onset of magnetic ordering. In the AFQ phase, the energy profile is largely independent of the polarization of the incident x-raysto within a scale factor, which is to be expected for a resonance that contains only a single order parameter (in this case, the quadrupole). Conversely, in the AFM + AFQ phase the structure in the energy profile for $\sigma$ and $\pi$ polarization of the incident x-rays contains quantitative differences in the ratios of all major features. This is indicative of a combination of magnetic and orbital scattering which interfere constructively or destructively as a function of energy. In the AFM+AFQ phase the intensity measured in the $\pi$ channel is generally larger than the intensity measured in the $\sigma$ channel, indicative of a strong magnetic contribution to the scattering, which is generally expected to be much larger for the M$_5$ edge than for the M$_4$ edge.

\section{Theoretical Framework}

It was demonstrated in the previous study \cite{us} that the unique shape of the energy dependence of the $\left( 0 0 \frac{1}{2} \right)$ 
reflection is well described by the splitting of the 3d core states. This splitting results in multiple interfering resonators and consequently adds structure to the energy profile in analogy to the unusual energy profile of the resonance at the M$_4$ edge in NpO$_2$ \cite{Np}. The addition of magnetism to the multipolar coulomb interaction in the combined AFM + AFQ phase further lifts the degeneracy of the 3$d$ shell with respect to the magnetic quantum number $\bar{M}$. Here $\bar{J}$ and $\bar{M}$ refer respectively to the total angular momentum and its projection for the 3$d$ hole state and $-\bar{J} \leq \bar{M} \leq \bar{J}$. The intra-atomic quadrupole interaction splits the different $|\bar{M}|$ levels, and the intra-atomic magnetic interaction lifts the remaining degeneracy in $\pm|\bar{M}|$. The relative oscillator amplitudes are determined by the 4f wavefunction, which is characterized in terms of multipoles.

Following the theory for resonance enhanced scattering of x-rays \cite{electronicproperties, paperone, papertwo, paperthree, paperfour}   and its extension to the AFQ phase of DyB$_2$C$_2$ \cite{us} we modify the equations for the scattered X-ray amplitude to include contributions from the time-odd multipoles allowed in the presence of the magnetic order that develops below T$_N$. These resonances due to odd-rank multipoles will interfere with the even rank ones either constructively or destructively. The X-ray scattering amplitude, the square of which is proportional to the X-ray intensity d$\sigma$/d$\Omega$, is formed from the product of two tensors \cite{electronicproperties}
\begin{equation}
f = \sum_{K} X^{K} F^{K} = \sum_{K,Q}(-1)^{Q} X^{K}_{-Q}F_{Q}^{K}
\end{equation}
with ${-K}\ \leq\ {Q}\ \leq {K}$. $F_{Q}^{K}$  describes the electronic response of the sample and $X_{-Q}^{K}$ is constructed from the polarization vectors of the incident and scattered X-ray beams, and is discussed explicitly in appendix A. $F_{Q}^{K}$ is expressed as
\begin{equation}
F_{Q}^{K} = (2K+1)^{1/2} \sum_{q} D^{K}_{Qq}(\alpha,\beta,\gamma)I_{q}^{K}
\end{equation}
where the Wigner D tensor $D_{Qq}^{K}(\alpha,\beta,\gamma)$ rotates $I_{q}^{K}$ (which refers to local coordinate axes), onto the coordinates of the experimental reference frame used for $X_{-Q}^{K}$ and $F_{Q}^{K}$ with Euler angles $\alpha,\beta,$ and $\gamma$. The Dy site symmetry in the magnetic phase of $2/m$ restricts $q$ to $q=\pm 1, \pm 2$. $K=2$ corresponds to quadrupoles (orbitals), while $K=1$ corresponds to magnetic dipoles, and the $K=0$ contribution corresponds to charge order, which is absent at the given Bragg reflection. $I_{q}^{K}$ describes the atomic resonance process which is commonly represented by a harmonic oscillator. In our previous work the standard theory has been extended so that $I_{q}^{K}$ is a sum of several oscillators created by the splitting of the core state. The amplitude $A_q^K(\bar{J},\bar{M})$ of each oscillator is labeled by the total angular momentum $\bar{J}=\frac{3}{2},\frac{5}{2}$ and magnetic quantum number $\bar{M}$ of the core hole created by the resonance process. 
\begin{equation}
I_{q}^{K} = \sum_{J,M}\frac{r_{\bar{J}} A_{q}^{K}(\bar{J},\bar{M})}{E-\Delta_{\bar{J}}-\epsilon^q(\bar{J},\bar{M})+i\Gamma_{\bar{J},|\bar{M}|}^{q}}
\end{equation}
\begin{equation}
\epsilon^q(\bar{J},\bar{M}) = {[3\bar{M}^{2}-\bar{J}(\bar{J}+1)]}Q_{\bar{J}}+g_{\bar{J}}\bar{M}\cdot \rm{H}_{\rm{Int}}
\end{equation}
where $E$ is the photon energy, $\Delta_{\bar{J}}$ is the difference in energy between the degenerate 3d$_{\bar{J}}$ shell and the empty 4f states and $\hbar/\Gamma_{\bar{J},\bar{M}}^{q}$ is the lifetime of the intermediate state. $\epsilon^q(\bar{J},\bar{M})$ is the energy shift of the core levels due to the intra-atomic quadrupole interaction $Q_{\bar{J}}$ and the intra-atomic magnetic interaction created by the unpaired 4f electrons H$_{\rm{Int}}$.

There are six resonant oscillators at the $M_5$ edge ($\bar{M}= \pm \frac{5}{2},\pm \frac{3}{2},\pm \frac{1}{2}$) and four at the $M_4$ edge ($\bar{M}= \pm \frac{3}{2},\pm \frac{1}{2}$). The oscillator amplitudes interfere and the branching ratio between the two edges is defined as a purely real mixing parameter $r_{\bar{J}}$. $A_q^K(\bar{J},\bar{M})$ is constructed from the structure factor of the chemical unit cell $\Psi_{q}^{x}$:

\begin{equation}
\fl A_q^K(\bar{J},\bar{M})=
(-1)^{\bar{J}-\bar{M}}\sum_{r}(2r+1)
\left(
\begin{array}{ccc}
\bar{J} & r & \bar{J} \\
-\bar{M} & 0 & \bar{M} \\
\end{array}
\right) 
\sum_{x} 
\left(
\begin{array}{ccc}
K & r & x \\
-q & 0 & q \\
\end{array}
\right) 
R^{K}(r,x)\Psi_{q}^{x}
\end{equation}
where the symbol in brackets is a 3j symbol \cite{juddbook} and $R^{K}(r,x)$ are reduced matrix elements, with $r=0,1,\ldots,2\bar{J}$, $x= | K - r |,\ldots,| K + r |$ and $q+x$ and $r+x$ are both even integers. The relevant reduced matrix elements $R^{K}(r,x)$ are listed in appendix C. 

The Dy site symmetry further dictates that $ A^2_2 = -A_{-2}^2$, and $ A_1^K = A_{-1}^K$, which follows from the properties of $\Psi_{q}^{x}$. The quantity $\Psi_{q}^{x}$ is constructed from the matrix elements of the multipole operator $\bf{T}$$_{q}^{x}$ combined with spatial phase factors and the symmetry components corresponding to the 4(c) site that the Dy occupies, discussed in detail in Appendix B and \cite{dymagneticdichroism}. For the (00$\frac{1}{2}$) reflection, this is
\begin{equation}
\Psi_{q}^{x}=\sum_{\bf d}e^{i{\bf \tau \cdot d}}\langle T_{q}^{x} \rangle_{\bf d}=(1-e^{\frac{i\pi q}{2}})\{\langle T_{q}^{x}\rangle-e^{\frac{i\pi q}{2}}\langle T_{-q}^{x}\rangle \}
\end{equation}
and 
\begin{equation}
\langle T_{q}^{x} \rangle = \langle \phi_0|$\bf{T}$_q^x|\phi_0\rangle
\end{equation}
Where $\tau$ is the Bragg wave-vector, $|\phi_0\rangle$ is the Dy 4f ground state wavefunction. Equations (5) to (7)  show that $A_1^1(\bar{J},\bar{M})$ and $A_1^2(\bar{J},\bar{M})$ are proportional to the dipole moment $\langle T_{1}^{1} \rangle$, octupole moment $\langle T_{1}^{3} \rangle$, and the triakontadipole moment $\langle T_{1}^{5} \rangle$ while $A_2^2(\bar{J},\bar{M})$ is proportional to the quadrupole moment $\langle T_{2}^{2} \rangle$, the hexadecapole moment $\langle T_{2}^{4} \rangle$, and the hexacontatetrapole moment $\langle T_{2}^{6} \rangle$ . 

The total intensity is represented by scattering from the dipole $\langle T_{1}^{1} \rangle$ and quadrupole  $\langle T_{2}^{2} \rangle$ of the 4f shell and reflects the same azimuthal angle dependence observed in the dipole transition at the L$_3$ edge \cite{massivepaper}. The dependence on the higher rank 4f multipoles arises from the splitting of the core hole states, which results in the $\bar{M}$ dependence of the amplitudes of the different harmonic oscillators. Correspondingly, these amplitudes are influenced by contributions from higher rank 4f multipoles which would otherwise cancel due to symmetry imposed by the 3$j$ symbols in (5). Therefore, the core hole splitting allows the measurement of higher rank multipoles through its spectral shape than would normally be expected for an E1 resonance. 

\section{Discussion}

\begin{figure}
\begin{centering}
\includegraphics[width=0.7\textwidth]{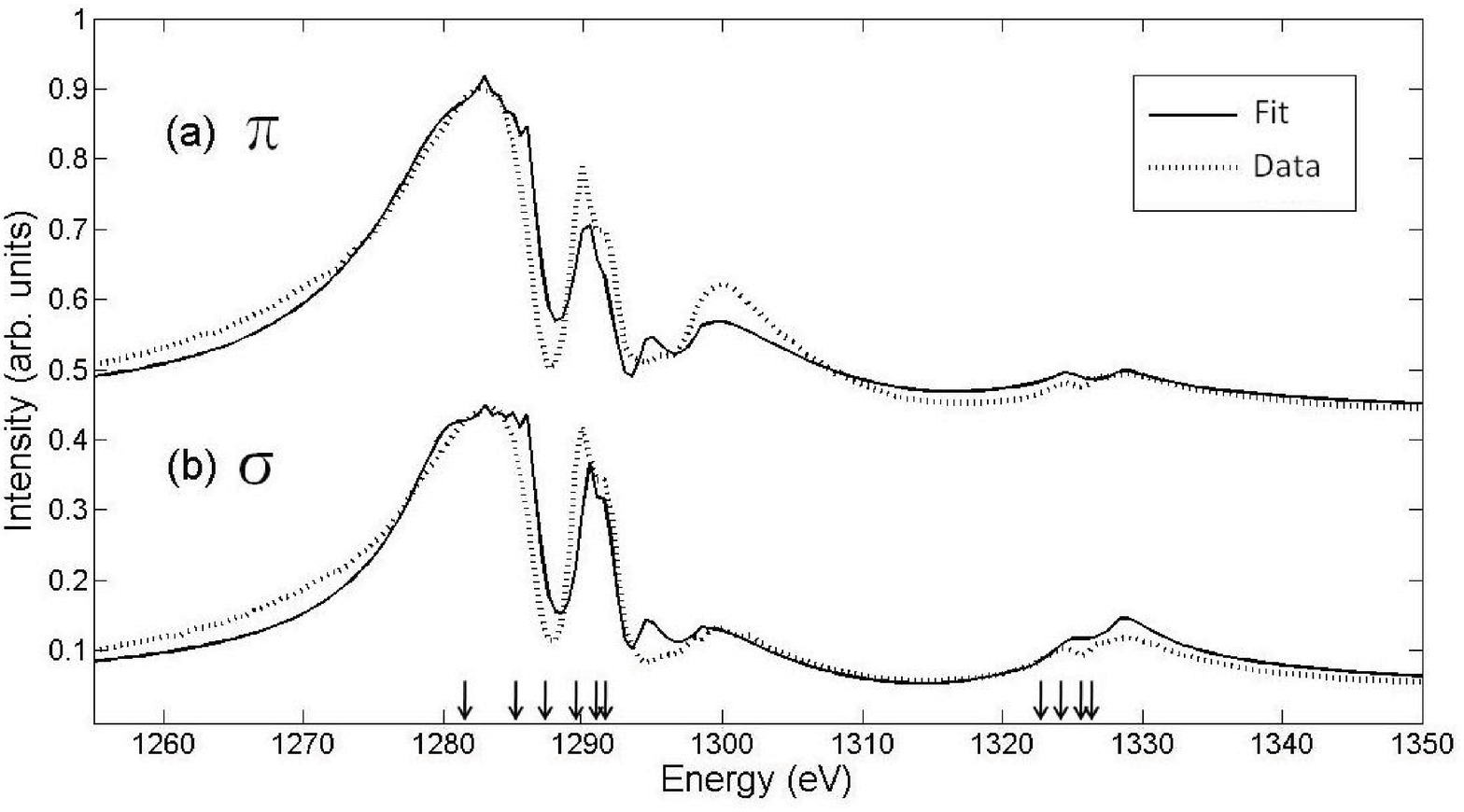}
\caption{\label{datafit} The best fit to the energy dependence for (a) $\sigma$, and (b) $\pi$ incident radiation in the AFM+AFQ phase of DyB$_2$C$_2$ at 12K}
\end{centering}
\end{figure}

The theory detailed in section 3 is used to fit the measured energy dependence for $\sigma$ and $\pi$ incident polarization via a nonlinear least squares optimization process. The best fit is shown in Figure \ref{datafit} with the oscillator positions indicated with arrows. The absorption correction is applied to the calculated spectra. In order to constrain certain features of the fitting procedure,  $\Gamma$$_{\bar{J},\pm \bar{M}}^{1}= a_s\Gamma_{\bar{J},\pm \bar{M}}^{2}$, where $a_s$ is a scaling parameter. The parameters obtained in the fit are shown in Table \ref{tab_fit}. $\Delta_{\bar{J}}$ typically depicts the onset of an absorption edge and the value of r$_{3/2} = 0.22$ calculated for the free ion \cite{vanderlaan} gives an excellent fit. We find that the quadrupole interaction Q$_{5/2}$ = $-0.42~eV$, and the internal magnetic field H$_{\rm{Int}}$ = $-0.98~eV$. H$_{\rm{Int}}$ is largely independent of the choice of wavefunctions and linewidths, and its order of magnitude is in line with that expected from atomic calculations of 3$d$-4$f$ exchange integrals \cite{vanderlaan}. Variations in $\langle T_{q}^{x} \rangle$ can be partially compensated by the line widths to yield a similar result. The linewidth for time-even contributions $\Gamma_{\bar{J},|\bar{M}|}^2$ increases progressively from $0.7~eV$ for $\Gamma_{\frac{5}{2},|\frac{1}{2}|}^2$ to $4.5~eV$ for $\Gamma_{\frac{3}{2},|\frac{3}{2}|}^2$ and is approximately a factor of 2.8 larger for time-odd contributions $\Gamma_{\bar{J},|\bar{M}|}^1$. $\Psi_{2}^{2}$ is normalized to 1 and the best fit for the values of $\Psi_q^x$ gives $\Psi_{2}^{4} = 0.22$, $\Psi_{2}^{6} = 0.07$, $\Psi_{1}^{1} = -0.34$, $\Psi_{1}^{3} = 0.77$, and  $\Psi_{1}^{5} = -0.78$. We find that the octupole and triakontadipole moments are the dominant time-odd contributions, rather than the dipole moment. This is in accord with observations that the elastic properties of DyB$_2$C$_2$ are similar to the related compound HoB$_2$C$_2$ for which a strong influence from octupolar order is inferred \cite{ultrasound}. We note in particular this is the first direct measurement of a triakontadipole moment.

It should be noted that the analysis in terms of a core-valence interaction is consistent with the multiplet structure interpretation. The M$_4$ and M$_5$ absorption structure is $\approx 10 eV$ and $\approx 5 eV$ wide respectively, which is mainly due to the strong 3$d$-4$f$ and 4$f$-4$f$ Coulomb and exchange interactions \cite{vanderlaan}. The strong electron correlation leads to non-diagonal matrix elements in the quantum numbers $\bar{L}$, $\bar{S}$, and $\bar{J}$ of the core hole. The levels of the same $\bar{M}$ connected by these non-diagonal elements excite partly coherently and this makes it difficult to separate the core hole states.  

We note that the empty states of the 4$f$ shell selected by the resonant diffraction process and their corresponding relative transition intensities are not necessarily the same as for an absorption process. Resonant diffraction is sensitive to the difference between electronic states and the conditions for nonvanishing intensity depend only on local symmetries (of the resonant atom) while absorption spectroscopy is sensitive to the average difference between electron states, and the conditions for nonvanishing terms must reflect the global symmetry (Neumann's principle) \cite{localsymmetry, anisotropicscattering}. For example, the magnetic moment of an antiferromagnet is not visible at the absorption edge, but resonant diffraction may yield large intensities proportional to the magnetic moment \cite{magnetismbook}. In the present theoretical framework the core hole is decoupled from the 4$f^{n+1}$ state and the effect of the 4$f$ multiplet structure is indirectly taken into account by the effective widths of the $\bar{M}$ distributions which exceed the intrinsic lifetime width and describe the data surprisingly well. In this way, much physical intuition regarding the associated atomic quantities of interest (i.e. the atomic tensors $\langle T_{q}^{x} \rangle$) is retained. This compares to recent numerical calculations \cite{JFR} which included a crystal field splitting of the $4f$ states and the full multiplet structure, where a similar magnitude of $Q_{\bar{J}}$ was found (although of different sign) but no information on the $\langle T_{q}^{x} \rangle$ could be extracted from such a calculation. This work further demonstrates that including the intra-atomic coulomb and magnetic interactions is a viable approach to account for the multiple spectral features and their broad distribution in energy. 
 
\begin{table}
\caption{\label{tab_fit} Fitted Parameters in the AFQ and AFM + AFQ phases of DyB$_2$C$_2$}
\centering
\begin{threeparttable}[c]
\begin{tabular}{ccc}
\\
Parameter & T = 18K (AFQ) & T = 12K (AFM + AFQ)\\
\hline
$\Delta_{5/2}$ & ${1287.7~eV}$ & ${1287.6~eV}$\\
$\Delta_{3/2}$ & ${1324.0~eV}$ & ${1323.9~eV}$\\
$\Psi_{2}^{2}$  & ${1^{\rm{a}}}$  & $1$\tnote{a} \\
$\Psi_{2}^{4}$  & ${-0.2}$ & ${0.22}$\\
$\Psi_{2}^{6}$  & ${0.3}$ & ${0.07}$\\
$\Psi_{1}^{1}$  & $-$ & ${-0.34}$\\
$\Psi_{1}^{3}$ & $-$& ${0.77}$\\
$\Psi_{1}^{5}$  & $-$ & ${-0.78}$\\
Q$_{\bar{J}}$ & ${-0.41~eV}$ & ${-0.42~eV}$ \\
$H_{Int}$ & ${-}$  & ${-0.98~eV}$\\
r$_J$ & ${0.22}$ & ${0.22}$ \\
$\Gamma_{\frac{5}{2},|\frac{1}{2}|}^{2}$ & ${0.8~eV}$ & ${0.7~eV}$ \\
$\Gamma_{\frac{5}{2},|\frac{3}{2}|}^{2}$ & ${2.7~eV}$ & ${2.8~eV}$ \\
$\Gamma_{\frac{5}{2},|\frac{5}{2}|}^{2}$ & ${5.3~eV}$ & ${3.2~eV}$ \\
$\Gamma_{\frac{3}{2},|\frac{1}{2}|}^{2}$ & ${1.9~eV}$ & ${2.2~eV}$ \\
$\Gamma_{\frac{3}{2},|\frac{3}{2}|}^{2}$ & ${5.4~eV}$ & ${4.5~eV}$ \\
$\Gamma_{\frac{5}{2},|\frac{1}{2}|}^{1}$ & ${-}$ & ${2~eV}$ \\
$\Gamma_{\frac{5}{2},|\frac{3}{2}|}^{1}$ & ${-}$ & ${7.9~eV}$ \\
$\Gamma_{\frac{5}{2},|\frac{5}{2}|}^{1}$ & ${-}$ & ${9.1~eV}$ \\
$\Gamma_{\frac{3}{2},|\frac{1}{2}|}^{1}$ & ${-}$ & ${6.2~eV}$ \\
$\Gamma_{\frac{3}{2},|\frac{3}{2}|}^{1}$ & ${-}$ & ${12.8~eV}$ \\
\hline
\end{tabular}
\begin{tablenotes}
\item [a] This value is fixed to 1
\end{tablenotes}
\end{threeparttable}
\end{table}

\section{Conclusion}

DyB$_2$C$_2$ exhibits an ordered phase with both antiferromagnetic (AFM) and antiferro-orbital (AFQ) order below T$_{\textrm{N}}$ = 15.3 K. Soft resonant x-ray Bragg diffraction experiments were performed at the Dy M$_{4,5}$ edges and the energy dependence of the space group forbidden$(00\frac{1}{2})$ reflection was measured in this low temperature AFM+AFQ phase. The Dy 4f multipole moments of rank 1 (dipole) to 6 (hexacontatetrapole) were measured and their magnitudes are determined to be between 7 and 78 \% of the the quadrupole (rank 2) moment. The dominant time-odd contributions are of rank 3 and 5, indicating a strong influence from octupolar and triakontadipole multipoles. The energy dependence of the $(00\frac{1}{2})$ reflection is modelled successfully by including an intra-atomic core-hole interaction parameterized by quadrupole and magnetic interactions. A pseudo-multiplet structure was also introduced, allowing for the intereference between oscillators that have different properties under time-reversal due to their having different excited-state lifetimes.

\section{Acknowledgements}

This work was supported by the Swiss National Science Foundation, and NCCR MaNEP. This work was partly performed at the Swiss Light Source of the Paul Scherrer Institute, Villigen, Switzerland. 

\appendix
\section{Explicit Relations for $X_{Q}^{K}$}

The quantity $X_{Q}^{K}$ appearing in Equation (1) is a spherical tensor constructed from components of the incident and reflected polarization vectors. It is described by Equation (66) in \cite{electronicproperties}, and the values appropriate to the current study are:
\begin{eqnarray}
\sigma'\sigma: \nonumber\\
X_{0}^{2} = \left( \frac{2}{3} \right) ^{1/2}\nonumber\\
\pi'\pi: \nonumber\\
X_{0}^{1} =\frac{i}{\sqrt{2}}\sin(2\theta),~X_{0}^{2} = -\frac{1}{\sqrt{6}}\cos(2\theta),~X_{2}^{2} =  \frac{1}{2} \nonumber\\
\pi'\sigma: \nonumber\\
X_{1}^{1} = -\frac{1}{2}e^{-i\theta},~X_{1}^{2} = -\frac{1}{2}e^{-i\theta} \nonumber\\
\sigma'\pi: \nonumber\\
X_{1}^{1} = \frac{1}{2}e^{i\theta},~X_{1}^{2} = -\frac{1}{2}e^{i\theta} \nonumber
\end{eqnarray}
Due to the inability to measure the polarization of the outgoing beam without severely attenuating the measured signal, the spectra in figures 1 and 2 are combinations of the rotated and unrotated channels. Specifically, $I_{\sigma}=I_{\sigma'\sigma}+I_{\pi'\sigma}$ and $I_{\pi}=I_{\pi'\pi}+I_{\sigma'\pi}$.

\section{Dy site symmetry and magnetic structure}

The structure factor appropriate to the Dy site symmetry and magnetic ordering

\begin{equation}
\Psi_{q}^{x}=(1-e^{\frac{i\pi q}{2}})\{\langle T_{q}^{x}\rangle-e^{\frac{i\pi q}{2}}\langle T_{-q}^{x}\rangle \}
\end{equation}
does not account for the slight rotations of the magnetic moments proposed in, for example, \cite{rbcmagnetism}. In the above case, we arrive at the following
\begin{eqnarray}
\Psi_{2}^{x}= 4i\langle T_{2}^{x} \rangle''\\
\Psi_{1}^{x}= -2i(\langle T_{1}^{x} \rangle'-\langle T_{1}^{x} \rangle'')
\end{eqnarray}
where $\langle \ldots \rangle'$ and $\langle \ldots \rangle''$ represent the real and imaginary parts respectively. If we include the proposed rotations of $\pm 9^\circ$ away from the site symmetry, we arrive at the following structure factor
\begin{equation}
\Psi_{q}^{x}=(1-e^{\frac{-9i\pi q}{180}}e^{\frac{i\pi q}{2}})\{\langle T_{q}^{x}\rangle-e^{\frac{9i\pi q}{180}}e^{\frac{i\pi q}{2}}\langle T_{-q}^{x}\rangle \}
\end{equation}
which gives 
\begin{eqnarray}
\Psi_{2}^{x}= -0.62i\langle T_{2}^{x} \rangle'' + 3.9i\langle T_{2}^{x} \rangle''\\
\Psi_{1}^{x}= -1.98i\langle T_{1}^{x} \rangle' + 1.69\langle T_{1}^{x} \rangle''
\end{eqnarray}

\section{Reduced matrix elements $R^{K}(r,x)$}

$R^{K}(r,x)$ is calculated for the Dy$^{3+}$ ground state of ${^6}H_{15/2}$ using

\begin{eqnarray}
R^K(r,x)=(2\bar{J}+1)(-1)^K(-1)^x\sqrt{2x+1}\sum_{a,b} ( {^6}H_{15/2}| W^{(a,b)x}| {^6}H_{15/2} ) \nonumber\\
\fl \times (2a+1)(2b+1) \sum_{y} (2y+1)
\left\{
\begin{array}{ccc}
K & r & x \\
a & b & y \\
\end{array}
\right\} 
\left\{
\begin{array}{ccc}
t & l & \bar{l} \\
t & l & \bar{l} \\
K & b & y \\
\end{array}
\right\} 
\left\{
\begin{array}{ccc}
1/2 & \bar{J} & \bar{l} \\
1/2 & \bar{J} & \bar{l} \\
a & r & y \\
\end{array}
\right\}
\end{eqnarray}
where the relation between W$^{(a,b)}$ and W$^{(a,b)x}$ is given by,
\begin{equation}
\fl ( \theta J| W^{(a,b)x}| \theta ' J' ) = \left\{ \frac{(2J+1)(2K+1)(2J'+1)}{(2a+1)(2b+1)}\right\}^{\frac{1}{2}}
\left\{
\begin{array}{ccc}
S & S' & a \\
L & L' & b \\
J & J' & K \\
\end{array}
\right\}
( \theta| W^{(a,b)}| \theta ' ) 
\end{equation}
$W^{(a,b)}$ is a unit double tensor in the defined in accord with Judd \cite{juddbook}, where a is the rank of the spin component and b is the rank of the orbital component. Its calculation is facilitated with the following equation:

\begin{eqnarray}
( l^n\alpha LS| W^{(a,b)}| l^n\alpha ' L'S') = n [L,L',b,S,S',a ] \sum_{\alpha,L,S}(-1)^{L+\bar{L}+b+\bar{S}+S+a+s+l}\nonumber\\
\fl \times \left\{
\begin{array}{ccc}
l & l & b \\
L & L' & \bar{L} \\
\end{array}
\right\}
\left\{
\begin{array}{ccc}
s & s & a \\
S & S' & \bar{S} \\
\end{array}
\right\}
( l^n \alpha L S\{| l^{n-1} \overline{\alpha L S})(l^{n-1} \overline{\alpha L S}|\}l^n \alpha ' L' S')
\end{eqnarray}
where $( l^n \alpha L S\{| l^{n-1} \overline{\alpha L S})$ is a coefficient of fractional parentage (CFP) and values for them are tabulated for almost all electron configurations of interest in Nielson \& Koster \cite{nielsonandcoster}, along with values of $(\theta||V(K)||\theta') = \left\{ \frac{1}{2}(2S+1)\right\}^{-1/2}(\theta||W^{0,K}||\theta')$. $s$ and $l$ are the spin and orbital angular momentum of the equivalent electrons coupled together in this process, and in the case of 4$f$ electrons have the values 3 and $\frac{1}{2}$ respectively. The results for the calculation of reduced matrix elements $R^K(r,x)$ for the Dy$^{3+}$ ion (which has $n_h = 5$ 4$f$ holes) in the ground state ${^6}H_{15/2}$ are tabulated in tables \ref{R_K2} and \ref{R_K1}. 

\begin{table}
\caption{\label{R_K1} Reduced matrix elements $R^1(r,x)$ for ${^6}H_{\frac{15}{2}}$ and n$_{\rm{h}}=5$ .}
\begin{center}
\begin{tabular}{r|ccccc}
\multicolumn{5}{c}{$K=1$, $\bar{J}=\frac{3}{2}$}\\
\hline
r  &0&1&2&3\\ \hline
$x$=0 & 0 & $\frac{4}{21}\sqrt{\frac{2}{15}}$ & 0 & 0 \\
1 & $\frac{4}{105}\sqrt{\frac{34}{15}}$ & 0 & $-\frac{8}{525}\sqrt{\frac{17}{15}}$ & 0\\
2 & 0 & $\frac{4}{21}\sqrt{\frac{17}{35}}$ & 0 & $-\frac{2}{49}\sqrt{\frac{34}{105}}$\\ 
3 & 0 & 0 & $-\frac{2}{175}\sqrt{\frac{646}{195}}$ & 0\\
4 & 0 & 0 & 0 & $-\frac{12}{245}\sqrt{\frac{1938}{455}}$\\
\hline
\end{tabular}

\begin{tabular}{r|ccccccc}
\multicolumn{7}{c}{$K=1$, $\bar{J}=\frac{5}{2}$}\\
\hline
r  &0&1&2&3&4&5\\ \hline
$x$=0 & 0 & $\frac{106}{63}\sqrt{\frac{1}{105}}$ & 0 & 0 & 0 & 0\\
1 & $\frac{2}{15}\sqrt{\frac{17}{5}}$ & 0 & $-\frac{6}{25}\sqrt{\frac{17}{35}}$ & 0 & 0 & 0\\
2 & 0 & $\frac{2}{441}\sqrt{\frac{34}{5}}$ & 0 & $\frac{1}{49}\sqrt{\frac{102}{35}}$ & 0 & 0 \\
3 & 0 & 0 & $-\frac{9}{175}\sqrt{\frac{646}{455}}$ & 0 & $\frac{22}{945}\sqrt{\frac{646}{455}}$ & 0\\
4 & 0 & 0 & 0 & $-\frac{62}{735}\sqrt{\frac{646}{1365}}$ & 0 & $-\frac{1}{1617}\sqrt{\frac{646}{195}}$\\
5 & 0 & 0 & 0 & 0 & $\frac{1}{27}\sqrt{\frac{646}{455}}$ & 0\\
6 & 0 & 0 & 0 & 0 & 0 & $\frac{2}{77}\sqrt{\frac{969}{35}}$\\
\hline
\end{tabular}
\label{table}
\end{center}
\end{table}

\begin{table}
\caption{\label{R_K2} Reduced matrix elements $R^2(r,x)$ for ${^6}H_{\frac{15}{2}}$ and n$_{\rm{h}}=5$ .}
\begin{center}
\begin{tabular}{r|cccc}
\multicolumn{5}{c}{$K=2$, $\bar{J}=\frac{3}{2}$ }\\
\hline
r &0&1&2&3\\ \hline
$x$=0 & 0 & 0 & $\frac{4}{105}\sqrt{\frac{2}{15}}$ & 0 \\
1 & 0 & $\frac{8}{2625}\sqrt{17}$ & 0 & $-\frac{4}{6125}\sqrt{\frac{34}{3}}$ \\
2 & $-\frac{2}{21}\sqrt{\frac{34}{35}}$ & 0 & $\frac{4}{735}\sqrt{17}$ & 0 \\
3 & 0 & $\frac{2}{875}\sqrt{\frac{646}{13}}$ & 0 & $-\frac{4}{2625}\sqrt{\frac{646}{273}}$ \\
4 & 0 & 0 & $\frac{36}{1225}\sqrt{\frac{969}{65}}$ & 0 \\
5 & 0 & 0 & 0 & $\frac{76}{735}\sqrt{\frac{323}{195}}$\\
\hline
\end{tabular}

\begin{tabular}{r|cccccc}
\multicolumn{7}{c}{$K=2$, $\bar{J}=\frac{5}{2}$ }\\
\hline
r &0&1&2&3&4&5\\ \hline
$x=0$ & 0 & 0 & $\frac{1}{45}\sqrt{\frac{14}{5}}$ &0 & 0 & 0\\
1 & 0 & $\frac{146}{7875}\sqrt{\frac{34}{7}}$ & 0 & $-\frac{71}{6125}\sqrt{\frac{34}{3}}$ & 0 & 0\\
2 & $-\frac{2}{21}\sqrt{\frac{17}{105}}$ & 0 & $-\frac{2}{735}\sqrt{\frac{17}{21}}$ & 0 &$\frac{2}{147}\sqrt{\frac{17}{35}}$ & 0 \\
3 & 0 & $\frac{6}{875}\sqrt{\frac{323}{91}}$ & 0 & $-\frac{58}{7875}\sqrt{\frac{646}{273}}$ & 0 & $\frac{2}{3465}\sqrt{\frac{4199}{105}}$ \\
4 & 0 & 0 & $\frac{66}{1225}\sqrt{\frac{323}{455}}$ & 0 & $-\frac{2}{735}\sqrt{\frac{646}{3003}}$ & 0 \\
5 & 0 & 0 & 0 & $\frac{22}{2205}\sqrt{\frac{323}{195}}$ & 0 & $\frac{46}{4095}\sqrt{\frac{323}{231}}$ \\
6 & 0 & 0 & 0 & 0 & $-\frac{1}{15}\sqrt{\frac{646}{385}}$ & 0 \\
7 & 0 & 0 & 0 & 0 & 0 & $-\frac{1}{1001}\sqrt{\frac{74290}{3}}$\\
\hline
\end{tabular}
\label{table}
\end{center}
\end{table}

\end{document}